\documentclass[%
 aip,
 amsmath,amssymb,
 reprint,%
]{revtex4-1}

\usepackage{graphicx}
\usepackage{dcolumn}

\usepackage[utf8]{inputenc}
\usepackage[T1]{fontenc}
\usepackage{mathptmx}
\usepackage{etoolbox}
\usepackage{txfonts}
\renewcommand*{\vec}[1]{\boldsymbol{#1}}
\usepackage{xcolor}

\makeatletter
\def\@email#1#2{%
 \endgroup
 \patchcmd{\titleblock@produce}
  {\frontmatter@RRAPformat}
  {\frontmatter@RRAPformat{\produce@RRAP{*#1\href{mailto:#2}{#2}}}\frontmatter@RRAPformat}
  {}{}
}%
\makeatother
\begin{document}


\title[]{
Enhanced MVA of Polarized Proton Beams via PW Laser-Driven Plasma Bubble
}

\author{Zhikun Zou}
\affiliation{%
 Department of Physics, Hubei University, Wuhan, 430062, China
}%

\author{Gan Guo}
\affiliation{%
 Department of Physics, Hubei University, Wuhan, 430062, China
}%

\author{Meng Wen}\email{wenmeng@hubu.edu.cn}
\affiliation{%
 Department of Physics, Hubei University, Wuhan, 430062, China
}%

\author{Bin Liu}
\email{liubin@glapa.cn}
\affiliation{%
Guangdong Institute of Laser Plasma Accelerator Technology, Guangzhou, China
}%

\author{Xue Yan}
\affiliation{%
School of Mechanical Engineering, Jiangsu University, Zhenjiang, 212013, China
}%

\author{Yangji\'e Liu}
\affiliation{%
 Department of Physics, Hubei University, Wuhan, 430062, China
}%

\author{Luling Jin}\email{jinluling@hubu.edu.cn}
\affiliation{%
 Department of Physics, Hubei University, Wuhan, 430062, China
}%

\date{14 February 2025}

\begin{abstract}
The significance of laser-driven polarized beam acceleration has been increasingly recognized in recent years.
We propose an efficient method for generating polarized proton beams from a pre-polarized hydrogen halide gas jet, utilizing magnetic vortex acceleration enhanced by a laser-driven plasma bubble.
When a petawatt laser pulse passes through a pre-polarized gas jet, a bubble-like ultra-nonlinear plasma wave is formed. As part of the wave particles, background protons are swept by the acceleration field of the bubble and oscillate significantly along the laser propagation axis.
Some of the pre-accelerated protons in the plasma wave are trapped by the acceleration field at the rear side of the target. This acceleration field is intensified by the transverse expansion of the laser-driven magnetic vortex, resulting in energetic polarized proton beams.
The spin of energetic protons is determined by their precession within the electromagnetic field, as described by the Thomas-Bargmann-Michel-Telegdi equation in analytical models and particle-in-cell simulations.
Multidimensional simulations reveal that monoenergetic proton beams with hundreds of MeV in energy, a beam charge of hundreds of pC, and a beam polarization of tens of percent can be produced at laser powers of several petawatts.
Laser-driven polarized proton beams offer promising potential for application in polarized beam colliders, where they can be utilized to investigate particle interactions and to explore the properties of matter under unique conditions.
\end{abstract}

\maketitle

\section{\label{sec:1}Introduction}

In recent years, high-powered laser facilities have become capable of delivering ultra-short pulses with an intense laser power surpassing 10 petawatts (PW)~\cite{Danson:hpl:2019,Li:LPR:2022} and a pulse duration of less than 20 femtoseconds (fs)~\cite{Li:18,Lureau_2020}.
The development of PW lasers is geared towards fundamental applications, including   fusion ignition systems and high-energy particle physics~\cite{Mourou:2006:RevModPhys.78.309,Mourou:2019:RevModPhys.91.030501}. 
Benefiting from the intense laser facilities, highly energetic beams of electrons, ions, and photons can be produced through laser-plasma interactions~\cite{Esarey:2009:RevModPhys.81.1229}. 
Ion acceleration driven by ultra-intense lasers in plasmas has been intensively investigated through theoretical models and experimental studies over the past decades~\cite{Macchi:2013:RevModPhys.85.751,Ma2021}.
Several efficient acceleration mechanisms in laser-solid interactions have been experimentally demonstrated, such as target normal sheath acceleration and radiation pressure acceleration~\cite{Wagner:2016:PhysRevLett.116.205002,Ma:2019:PhysRevLett.122.014803,Kim::POP:2016,Wang:2020:PhysRevLett.125.034801}.
An experimental record of generating 150 MeV proton beams from a solid foil has been reported recently~\cite{Ziegler:NPhy:2024,Bin2024}. 
Furthermore, ion acceleration at high repetition rates naturally occurs in hydrodynamic flows with near-critical density~\cite{Sylla:2012:RSI,Prencipe:HPLSE:2017,Ospina:2024:PhysRevResearch.6.023268}, where ions can be accelerated through various mechanisms like bubble acceleration (BA)~\cite{Shen:2007:PhysRevE.76.055402,Zhang:NJP:2014}, shock acceleration (SA)~\cite{Palmer:2011:PhysRevLett.106.014801,Haberberger:NPhy:2012}, and magnetic vortex acceleration (MVA)~\cite{Nakamura:2010:PhysRevLett.105.135002,Park:2019:POP,Tazes:SR:2024}.

The energetic polarized particle beam plays a pivotal role in high energy physics and nuclear physics. The collision of polarized particle beams provides a unique opportunity to study the spin structure of hadrons~\cite{Roser:2003:AIPCP,Burkardt:RPP:2010,Zhang:2019}. 
Polarized particle beams are produced in traditional accelerators, which are expensive and huge. For instance, the proton accelerators of the RHIC and JLEIC consist of an ion linac injector with an energy of approximately 200 MeV and energy boosters with hundreds of GeVs~\cite{Alekseev:2003,MartinezMarin:2019odq,Mustapha_2020}. With the rapid development of PW laser facilities, compact and economical laser plasma accelerators have come to the rescue.
Polarized particles can be generated by radiation polarization of a laser-accelerated beam~\cite{Li:2019:PhysRevLett.122.154801,Samsonov:2023:MRE,Tang:2024:MRE} or by laser acceleration in a pre-polarized target~\cite{Wen:2019:PhysRevLett.122.214801,Gong:2023:MRE}. Notably, laser-driven proton acceleration only works with the latter method due to the significantly larger mass-to-charge ratio of protons compared to electrons~\cite{huetzenHPLSE19,Jin:2020:PhysRevE.102.011201}.
Recently, the target with polarized protons can be provided by the dense polarized atomic hydrogen gas jets~\cite{Sofikitis:2018:PhysRevLett.121.083001,Spiliotis2021}. 
The literatures on laser-driven polarized proton acceleration using these dense gas targets primarily focused on feasibility verification, with studies showing one hundred MeV polarized proton beams generate with PW lasers via the MVA                                                                                                                                                                                                                                                                                                                                                                                                                                                                                                                                                                                                                                                                                                                                                                                                                            scheme~\cite{Jin:2020:PhysRevE.102.011201,Reichwein:2022:PhysRevAccelBeams.25.081001} or boosted SA~\cite{Yan:PPCF:2023}. 
The BA scheme has been demonstrated successful in producing GeV polarized proton beams with a 200 PW laser~\cite{Li:2021:PhysRevE.104.015216}. It becomes challenging with the currently available laser power due to the self-injection threshold~\cite{Liu:2016:PhysRevAccelBeams.19.073401,Liu:POP:2018}.
Current intense laser facilities provide PW pulses, while the hundred PW laser is still under construction~\cite{Danson:hpl:2019,Li:LPR:2022}.
Considering the requirements for polarized colliders~\cite{Roser:2003:AIPCP,Burkardt:RPP:2010,Zhang:2019}, it is worthwhile to pursue increasing the acceleration efficiency of polarized protons using a PW laser.

In order to enhance the laser driven ion acceleration in a gas jet, useful techniques has been applied experimentally, including the tightly focusing of PW pulses near diffraction-limited~\cite{Yoon:21:Optica,Wang:2021:PhysRevX.11.021049}
 and the high density gas jet production with a hydrodynamic shock wave excited by a low-energy laser pulse~\cite{Marques:2023:MRE,Seemann:2024:PhysRevLett.133.025001}. 
Propagation of such intense pulses in the dense hydrodynamic jets may excite bubble-like ultra-nonlinear plasma waves, where the intense oscillation of background ions becomes a part of the plasma wave. The condition of self-injection of the background ions in the plasma wave is relatively challenging for PW lasers.
Nevertheless, the background ions in the plasma wave acquire significant energy, which could be expected to boost the further injection and acceleration in MVA at the rear side of the target.

In this paper, we demonstrate the generation of monoenergetic polarized proton beams with energies in the hundreds of MeV range from a PW laser in a pre-polarized hydrogen chloride gas jet within an enhanced MVA. When the intense laser pulse penetrates the target, atoms with pre-polarized nuclei are ionized and a nonlinear plasma wave is excited by a ponderomotive force. The motion of background electrons and ions induces strong electromagnetic fields in the laser plasma wave. The fields consist of the charge separation field and a vortex magnetic field, where the proton energy enhances and its spin precesses. Before the laser pulse leaves the target, protons are accelerated by the laser driven plasma wave, in radial and forward directions, respectively. Dependence of proton precession on laser plasma parameters are analyzed.  When the laser pulse leaves the target, the laser driven magnetic vortex  introduces a stationary electric field at the rear boundary. Protons from the density down ramp still oscillate longitudinally in the plasma wave, which may be trapped and  further accelerated in the magnetic vortex enhanced stationary acceleration field. Our particle-in-cell (PIC) simulations indicate that monoenergetic polarized proton beams with energies of several hundred MeV can be generated using PW lasers.

\section{Simulation setup}

\begin{figure}[t]
\includegraphics[width=0.45\textwidth]{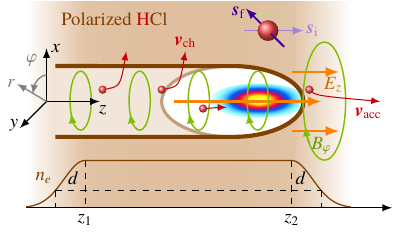}
\caption{\label{fig:1} The schematic diagram showing the drive laser irradiating onto a gas jet of polarized HCl. A plasma channel is formed behind the laser bubble.  Some background protons are pushed longitudinally by $E_z$, and protons with $\vec v_\mathrm{ch}$ are accelerated radially by the transverse field in the plasma channel. 
 Meanwhile, the spin of protons precesses in the magnetic vortex according to Eq.\eqref{tbmteq}. Transverse expansion of the magnetic vortex enhance the acceleration field at the rear boundary of the target. The forward moving protons are trapped and accelerated to several hundred MeV.}
\end{figure}

The driving laser is a tightly focused Gaussian pulse, with a focal radius of $w_0 = 3 \lambda$, a laser amplitude of $a_0=120$, a pulse duration of $T_L = 10 \tau$, and linear polarization along the $y$-axis, where $\tau = \lambda /c$ denotes the laser cycle and $\lambda = 800$~nm is the laser wavelength.
The corresponding laser power  
$\mathcal{P}_L= (\pi^2/4) \mathcal{P}_0 a_0^2 w_0^2/\lambda^2$
attains 2.78~PW, where $\mathcal{P}_0 = 4\pi \varepsilon_0 m_e^2 c^5/e^2 \approx 8.7$~GW is the natural
relativistic power unit, with $\varepsilon_0$ the vacuum permittivity, $m_e$ and $-e$ the electron mass and charge. 
The schematic diagram illustrating the proton acceleration and spin depolarization in the laser-driven plasma field is presented in Fig.~\ref{fig:1}. A PW laser pulse irradiates a HCl gas target containing pre-polarized hydrogen atoms with relatively high density. The atoms are ionized rapidly with the spin of their nuclei conserved.
Such an ultra-intense PW pulse can penetrate into overdense plasma with density $n_e\lesssim a_0 n_c/2$~\cite{Liu2024}, where $n_{c}=\varepsilon_0 m_e \omega_L^2/e^2$ is the  plasma critical density and $\omega_L$ is the laser frequency. 
A plasma channel is formed when the laser power exceeds $2\mathcal{P}_0$~\cite{Borisov:1992:PhysRevLett.68.2309,Shen:2022:PhysRevApplied.18.064091}. 
The channel radius depends on the laser power and the plasma density according to~\cite{Park:2019:POP,Gu_Bulanov_2021_HPLSE} 
\begin{align}
\label{eq:rch}
\dfrac{R_\mathrm{ch}  }{\lambda}\approx \left(\dfrac{a_0}{2\pi^2 \sqrt{K} } \dfrac{w_0}{ \lambda}\dfrac{n_{c}}{n_e}\right)^{1/3}  \,.
\end{align}
 Inside the plasma channel, the magnetic field produced by the laser driven hot electrons and returned current depends linearly to the channel radius~\cite{Pukhov:1996:PhysRevLett.76.3975}. It is therefore estimated as
 \begin{align}
 \label{eq:bphi}
 B_\varphi \approx \left(\dfrac{\pi a_0}{ \sqrt{K} } \dfrac{w_0}{ \lambda}\right)^{1/3} \left(\dfrac{n_e}{n_{c}}\right)^{2/3} \dfrac{m_e \omega_L}{e} \,.
\end{align}
This magnetic vortex reaches $10^5$~Tesla, which
results significant proton spin precession.

Two-dimensional PIC simulations are performed using the EPOCH code, which has been extended to include spin effects~\cite{Arber:epoch:2015hc}. 
We use a moving window of size $180\lambda \times 80 \lambda$ represented by the grid of $3600 \times 1600$, with 4 pseudo-particles per cell for protons, 1 pseudo-particle per cell for chlorine nuclei, and 4 pseudo-particles per cell for electrons.
Additionally, a time-step according to the Courant–Friedrichs–Lewy condition with a Courant number of 0.95 is utilized in the simulations. 
The laser pulse is designed to focus onto the left boundary of the gas target at $z_1=50\lambda$. The target consists of a uniform plateau of $n_{e} = 0.54 n_c$ in the region $z_1\leq z \leq z_2$, and edges $n_{e}\exp\left[-(z-z_j)^2/d^2\right]$ on the up-ramp side  $z<z_1$ and the down-ramp side  $z>z_2$, where $j=1,\,2$, $d = 5 \lambda$ and $z_2= 250\lambda$. The corresponding densities of protons and chloride ions in the plasma are denoted as $n_p= n_{Cl}=n_e/18$. 
The target thickness is optimized laser depletion length $\sim a_0 T_L K n_c/n_e$~\cite{Bulanov:2010:10.1063/1.3372840}.
The plasma target consists of electrons and nuclei derived dissociated from pre-polarized HCl molecules~\cite{Spiliotis2021}, with all protons initially polarized along the axis of laser propagation $\vec s_\mathrm{i} = \vec{\mathrm{e}}_z$.
The proton spin precession in the PIC code is described by the Thomas-Bargmann-Michel-Telegdi (TBMT) equation 
\begin{align}
\label{tbmteq}
\dfrac{d\vec s}{dt} = \dfrac{e}{m_p} \vec s \times \left[\left(G+\dfrac{1}{\gamma}\right)\vec B - \dfrac{G\gamma}{\gamma+1}\dfrac{\vec v \cdot \vec B}{c^2}\vec v \right. \nonumber\\
\left.- \left(G + \dfrac{1}{\gamma+1}\right)\dfrac{\vec v}{c^2} \times \vec E\right],
\end{align} 
where $G \approx 1.79$ represents  the anomalous magnetic moment of proton, $\gamma$ is the relativistic Lorentz factor and $m_p$ is the proton mass. In the unrelativistic limit $v\ll c$, the precession frequency is determined by the first term of Eq.~\eqref{tbmteq}, which depends linearly on the magnetic field. In the magnetic vortex of the laser driven plasma channel, we have 
\begin{align}
\label{eq:omega}
\Omega= \left(G+1\right) \dfrac{m_e}{m_p} \left(\dfrac{\pi a_0 }{2\sqrt{K} } \dfrac{w_0}{ \lambda}\right)^{1/3} \left(\dfrac{n_e}{n_c}\right)^{2/3} \omega_L\,.
\end{align}

\section{Proton dynamics in gas jet}

\begin{figure}[t]
\includegraphics[width=0.45\textwidth]{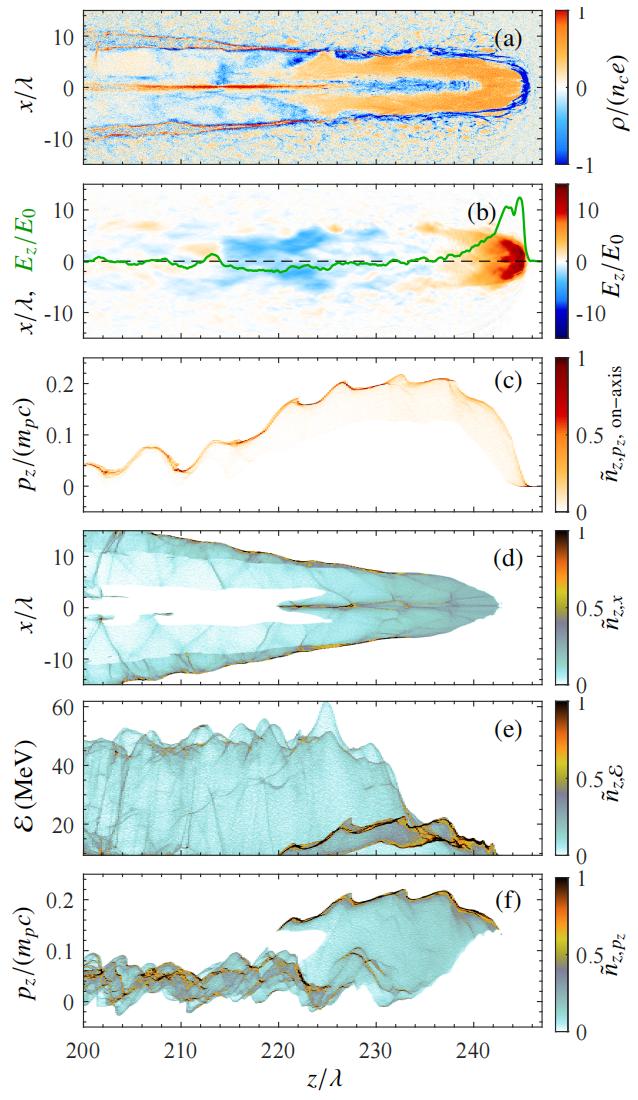}
\caption{\label{fig:2} 
Snapshots are presented of (a) the plasma charge density $\rho$ and (b) the longitudinal electric field $E_z$, along with density plots of on axis protons (with $\left|x\right|<2\lambda$) in space along the $z$ axis and longitudinal momentum $p_z$ (c) at $t=250\tau$.
Density plots of energetic protons (with energy larger than $\mathcal{E}_\mathrm{min}=9.4$~MeV) in space, specifically (d) along the $z$ and $x$ axes, (e) along the $z$ axis and energy $\mathcal E$, and (f) along the $z$ axis and longitudinal velocity $p_z$ at $t=250\tau$. The charge density of the plasma in (a) is comprised of the charge densities of electrons, protons, and chlorine nuclei. The solid green line in (b) depicts the longitudinal electric field along the laser propagation axis with $x=y=0$.
} 
\end{figure}

When an intense short laser pulse propagates through underdense plasma, a laser-driven bubble is formed, where electrons are completely blown out ponderomotively. Figure~\ref{fig:2}(a) shows an electron bubble with positive charges surrounding the drive laser within the plasma.
The density of the positive charges is approximately $\rho \approx n_{e0} e$.
In the ultra-nonlinear scheme driven by a PW laser, the electron bubble is adjoined by a relativistic plasma channel with the same radius. 
In the adjoined  bubble-channel structure, the electric field is strong enough to accelerate background ions~\cite{Shen:2007:PhysRevE.76.055402,Singh:SR:2020}.
The simulations are initialized as the intensity peak of the laser pulse reaches the position $z=0$ at time $t=0$.  
Figure~\ref{fig:2}(a) shows the charge density distribution at time $t=250\tau$, when the laser front arrives at a position $z_\mathrm{f} \approx 245 \lambda$. 
Electrons along the laser propagating axis are expelled forward to $z_\mathrm{f}$ and radially within the range $z_\mathrm{f}- c T_L\leq z\leq z_\mathrm{f}$, thereby forming the sheath of the front half of the bubble.
Under the space charge effect, background electrons are drawn back towards the propagating axis, which enclose the bubble at $z_\mathrm{t} \approx 220 \lambda$.
Furthermore, Eq.~\eqref{eq:rch} gives the plasma channel radius $R_\mathrm{ch}\approx 4.76 \lambda$ by taking the geometrical factor of 2D simulation $K \approx 1/10$~\cite{Bulanov:2010:10.1063/1.3372840}, which is coincide with the density boundary of the plasma channel behind the bubble in the region $z<z_\mathrm{t}$, as shown in Fig.~\ref{fig:2}(a).
Beside the sheath of the bubble and the plasma channel, a filament composed of dense electrons and ions appears on the propagation axis at $x\approx 0$ due to the motion of ions~\cite{Popov:2010:PhysRevLett.105.195002,Ji:2014:PhysRevLett.112.145003}. This filament exhibits distinct characteristics: inside the bubble, it mainly consists of trapped electrons due to the different charge-mass ratios of electrons from ions, shown as the negative charge density; whereas behind the bubble, the positive charges are dominated by background ions that are pulled along the propagation axis by the electron filament.

The charge density distribution results in cylindrically symmetric charge separation fields, $E_z$ and $E_\mathrm{ch}$, indicating the longitudinal electric field of the bubble and the radial electric field of the plasma channel, respectively.
 Figure~\ref{fig:2}(b) shows that the longitudinal component of the electric field in the bubble has a large amplitude at the front of a plasma bubble $z=z_\mathrm{f}$. The peak field strength is estimated as~\cite{Liu_2020_PPCF,Goethel_2022_PPCF}
\begin{align}
\label{eq:ezm}
E_{z,\mathrm{max}}\approx \dfrac{\pi}{2^{1.5}}\left(1+\dfrac{v_\mathrm{f}}{c}\right)\sqrt{\dfrac{a_0n_e}{n_c}}E_0\,,
\end{align}
with $v_\mathrm{f} =  c \left/\sqrt{1+\pi^2 n_e/a_0}\right.$ the velocity of the dense electron layer at the front of the laser bubble, and $E_0 = m\omega c/e$ the unit of the electric field.
Beside the charge separation and electric field, the longitudinal motion of background protons is another character of a ultra-nonlinear plasma wave~\cite{Liu:2016:PhysRevAccelBeams.19.073401,Liu:POP:2018}.
The momenta of protons on propagating axis with $r<2\lambda$ are shown in Fig.~\ref{fig:2}(c), where the maximum momentum is found in the middle of the bubble.
Considering Eq~\eqref{eq:ezm}, the protons oscillate in the plasma wave with the maximum momentum
\cite{Liu:2022:PhysRevLett.129.274801}
\begin{align}
\label{eq:pzm}
p_{z,\mathrm{max}} = \dfrac{\pi^2}{16}\left(1+\dfrac{c}{v_\mathrm{f}}\right)^2a_0 m_e v_\mathrm{f}\,.
\end{align}

\begin{figure}[t]
\includegraphics[width=0.45\textwidth]{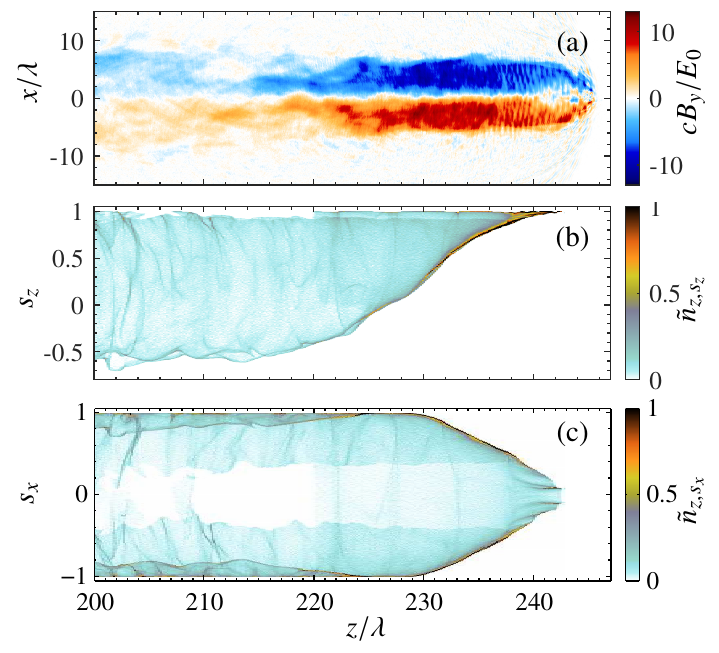}
\caption{\label{fig:3} (a) Snapshot of azimuthal magnetic field and density plots of energetic protons (with energy $\mathcal{E}>9.4$ MeV)  in space with (b) $z$ axis and longitudinal spin component $s_z$, and (c) $z$ axis and transverse spin component $s_x$ at $t=250\lambda/c$.  The $y$-component of the magnetic field $B_y$ in (c) presenting the azimuthal magnetic field $B_\varphi$ at the plane of $y=0$. }
\end{figure}

 Density plot of energetic protons with energy $\mathcal{E} > 9.4$~MeV  in space of $z$ axis and $x$ axis is shown in Fig. \ref{fig:2}(d). There are two groups of energetic protons on and off propagating axis, respectively. Density plots of these energetic protons at $t=250\tau$ in space of $z$ axis and energy $\mathcal E$, as well as in space of $z$ axis and longitudinal momentum $p_z$ are shown in Figs.~\ref{fig:2} (e) and (f), respectively.  
The on-axis group with $r< 2\lambda$ in Fig. \ref{fig:2}(d) occupies the same region of the bubble $z_\mathrm{t}<z<z_\mathrm{f}$. These protons are pushed forward when the front half of the bubble passes by. Corresponding dense protons can be found within the bubble region of Figs.~\ref{fig:2}(e) and (f), with $\mathcal{E}\lesssim 20$~MeV and $p_z \gtrsim 0.15 m_p c$. 
The momentum given by Eq.~\eqref{eq:pzm} is about $0.16m_p c$, consistent qualitatively with Figs.~\ref{fig:2}(c) and (f).
Here, one should note that $v_{z,\mathrm{max}} \ll v_{\mathrm{f}}$, with $v_{z,\mathrm{max}} = p_{z,\mathrm{max}}/(\gamma m_p c)$. This indicates that the on-axis group has not been trapped by the bubble, which is different from the self-injection case of BA~\cite{Liu:2016:PhysRevAccelBeams.19.073401,Liu:POP:2018,Li:2021:PhysRevE.104.015216}.
The on-axis protons are accelerated in the bubble, acquiring forward velocity in the front half of the bubble, and decelerated in the rear half.
 Besides, the spin of protons precesses in the plasma channel due to the magnetic field given according Eqs.~\eqref{eq:bphi} and \eqref{eq:omega}. The magnetic vortex at $t=250\tau$ is shown in Fig.~\ref{fig:3}(a). Furthermore, distributions of energetic protons on the spin components $s_z$ and $s_x$ are shown in in Figs.~\ref{fig:3}(b) and (c), respectively. 
 The distribution of protons on $s_x$ is symmetric, since the proton spin precesses in the cylindrically symmetric magnetic vortex.
 Since the singularity of the magnetic vortex is located at $\sqrt{x^2+y^2}=0$, the spin precession of on-axis BA protons is restricted. 
 Specifically, the spin distribution of the protons near the bubble front $z\rightarrow 240 \lambda$ in Figs. \ref{fig:3}(b) and (c), where $s_z \rightarrow 1$ and $s_x \rightarrow 0$, i. e., $\vec s \rightarrow \vec s_i$.

 Notably, the radial acceleration of background protons in the bubble-channel seems also promising inside the gas jet.
The off axis group of protons with $x\gtrsim R_\mathrm{ch}$   in Fig. \ref{fig:2}(d) is mainly accelerated by  $E_\mathrm{ch}$ and moves radially in the region $z \leq (z_\mathrm{f}+z_\mathrm{t})/2 $. The corresponding energy shown  in Fig. \ref{fig:2}(e) is consistent with the proton energy accelerated by the strong electrostatic sheath field of the channel $\mathcal{E}_\mathrm{ch} = a_0 mc^2$~\cite{Pukhov:1999:POP,Singh:SR:2020}. In other words, the SA protons are accelerated to velocity $v_\mathrm{ch} \approx \sqrt{2\mathcal{E}_\mathrm{ch}/m_p}$, during the time duration $t_\mathrm{ch}\approx 2R_\mathrm{ch}/v_\mathrm{ch}$. The time duration is then determined by the laser intensity and plasma density as
\begin{align}
t_\mathrm{ch} =  \dfrac{\lambda}{c} \left(\dfrac{2 }{K a_0 } \right)^{1/6} \left( \dfrac{w_0}{ \pi^2 \lambda}\dfrac{n_c}{n_e}\right)^{1/3} \sqrt{\dfrac{m_p}{m_e}} .
\end{align} 
 Moreover, the proton spin precesses in the magnetic vortex of Eq.~\eqref{eq:bphi}. It is roughly $8 E_0/c$ with our applied parameters, comparable to the field shown in Fig.~\ref{fig:3}(a). Considering the precession frequency of Eq.~\eqref{eq:omega}, the precession angle is obtained
 \begin{align}
\Theta = (G+1) \left(2a_0 \right)^{1/6} \left( \dfrac{2 w_0}{ \lambda}\right)^{2/3}  \left( \dfrac{\pi^2}{  K}\dfrac{n_e}{n_c}\right)^{1/3} \sqrt{\dfrac{m_e}{m_p}}  .
\end{align}
Since we apply the PW laser and a near critical dense gas jet, the precession angle of the SA protons $\Theta$ becomes roughly 2~radians. The corresponding final longitudinal spin component $s_z=\cos \Theta$ approaches -0.43, in agreement with the minimum of $s_z$ in Fig. \ref{fig:3}(b).

\section{Fields in the rear side of the target}

\begin{figure}[t]
\includegraphics[width=0.45\textwidth]{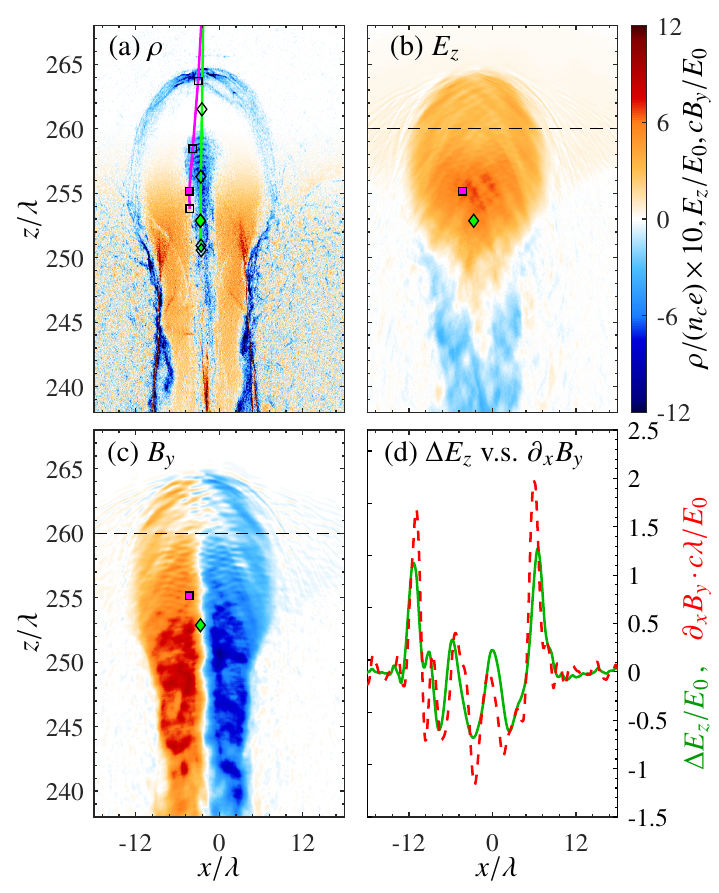}
\caption{\label{fig:4} Snapshots of (a) plasma charge density, (b) longitudinal electric field and (c) azimuthal magnetic field at $t=270\tau$. In (d), we compare the electric field increment in one laser cycle $\Delta E_z = \left(E_x|_{t=270.5\tau} - E_x|_{t=269.5\tau} \right) $ with the radial gradient of the azimuthal magnetic field $\partial_r B_\varphi|_{y=0} = \partial_x B_y$ at $z=270\lambda$, which are plotted with solid green and dashed red lines, respectively. In (a), the light green and magenta lines show the trajectories of two traced  protons, with diamond and cube marks indicating positions of the two protons along their trajectories. The marks field with green and magenta in (a)-(c) indicate the protons positions at $t=270\tau$.}
\end{figure}

When the drive pulse passes through the rear surface of the gas jet, a stationary acceleration field in the magnitude of Eq.~\eqref{eq:ezm} appears in the region behind the surface $z>z_2$. 
The charge density,  acceleration field and azimuthal magnetic field at $t=270\tau$ are  shown in Figs.~\ref{fig:4}(a)-(c), respectively.
It is found that the bubble shell expands transversely in the region $z>z_2$, with the companion of transverse expansion of $E_z$ and $B_\varphi$.
In order to show the expounding property along transverse direction, we set horizontal coordinate with $x$-axis in Fig.~\ref{fig:4}. Figure~\ref{fig:4}(a) shows half of the bubble leaves the flattop density of the gas jet at $t=270 \tau$. The bubble radius expands to $ 10\lambda$ at $z>z_2$, in comparison with the radius $R_\mathrm{ch} \approx 5\lambda$ inside the gas jet with $z<z_2$. 
 The acceleration field $E_z$ in Fig.~\ref{fig:4}(b) shows a static lower boundary at $z_2$, while its upper boundary moves with the laser-driven electron layer. It is worth noting that the acceleration field sustains $E_{z} \gtrsim E_0$, as a result of electric field enhancement via the gradient of the magnetic vortex
\begin{align}\label{ext}
\partial_t E_z = c^2 \partial_r B_\varphi  \,.
\end{align}  
The increment rate of electric field at $z=270\lambda$ is plotted with the solid green curve in  Fig.~\ref{fig:4}(d). It is consistent qualitatively with the radial gradient of the magnetic vortex shown with the dashed red curve. Proton acceleration in this stationary magnetic vortex enhanced acceleration field is thus called MVA~\cite{Nakamura:2010:PhysRevLett.105.135002}.  The axis of the magnetic vortex deviate a little from the initial laser propagating axis, as an effect of the hydrodynamic instability of the plasma~\cite{Huang:2017:PhysRevE.95.043207}.

\begin{figure}[t]
\includegraphics[width=0.45\textwidth]{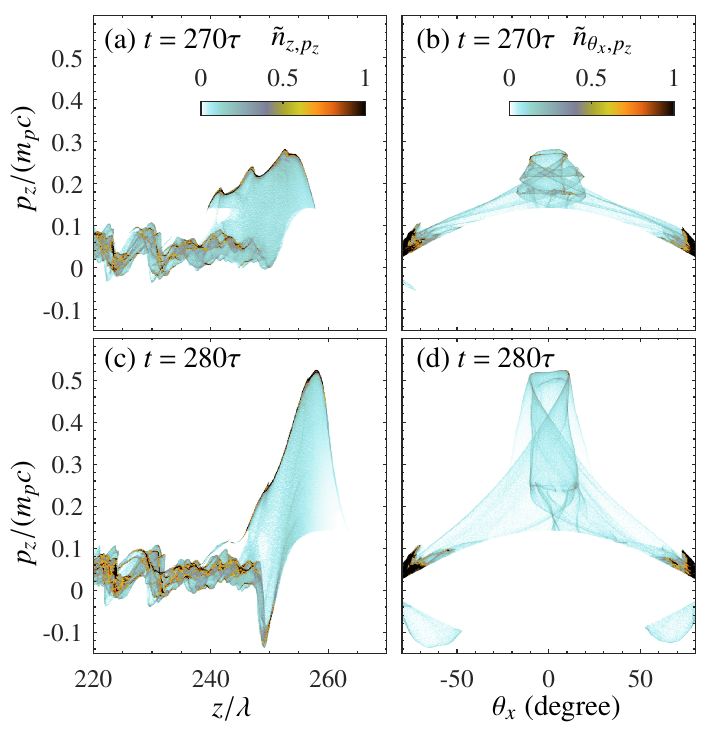}
\caption{\label{fig:5} Density plots of energetic protons in space along the longitudinal velocity $p_z$ with the $z$ axis (a) and the spread angle \mbox{$\theta_x = \tan^{-1}\left(p_x/p_z\right)$} (b) at $t=270\tau$. Same density plots at $t=280\tau$ are shown in (c) and (d).
} 
\end{figure}

Energy enhancement of protons via MVA is the second stage acceleration for the on axis protons near the rear boundary of the target. They are accelerated firstly when they are swept by the acceleration field of the bubble, described by Eqs.~\eqref{eq:ezm} and \eqref{eq:pzm}. The momentum distribution of the energetic protons along the propagating axis before they leave the gas target are shown in Fig.~\ref{fig:2}(f) at $t=250\tau$. 
When the bubble passes the rear boundary at $z_2$, the pre-acceleated protons in the plasma wave are then trapped by the stationary field of MVA, and accelerated to much higher energies.
When the accelerated protons pass $z_2$, the longitudinal momentum component reaches $0.28m_p c$ at $t=270\tau$ and $0.52m_p c$ at $t=280\tau$, as depicted in Figs.~\ref{fig:5}(a) and (c), respectively.
Figures~\ref{fig:5}(b) and (d) present the momentum distributions of energetic protons along the spread angle $\theta_x$, indicating that only protons with small spread are accelerated forward. The accelerated beam comprises protons within a narrow spread angle, specifically $\theta_x \lesssim 10^\circ$.

\begin{figure}[t]
\includegraphics[width=0.45\textwidth]{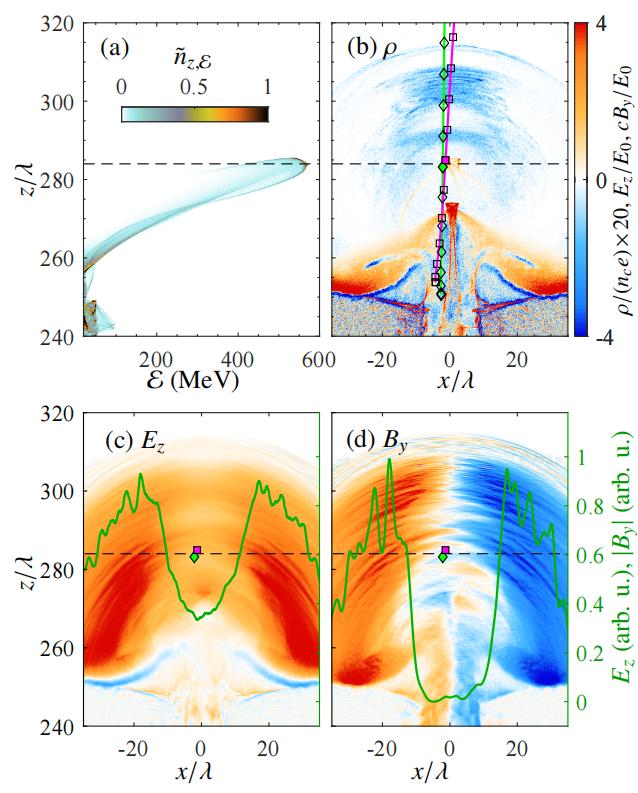}
\caption{\label{fig:6} Snapshots of (a) the phase space distribution of energetic protons,(b) the plasma charge density,  (c) the longitudinal electric field and (d) the azimuthal magnetic field at $t=320\tau$. The dashed black lines indicate the position $z = 284 \lambda$, where the maximum energy of protons appears. The corresponding longitudinal electric field and the azimuthal magnetic field at $z = 284 \lambda$ are given by solid dark green lines in (c) and (d), respectively. The trajectories and positions of two traced protons shown with light green and magenta lines and marks in (b)-(d) are same as those in Fig.~\ref{fig:4}. 
} 
\end{figure}

  Figure~\ref{fig:6}(a) shows protons are accelerated to 550~MeV by $t=320\tau$, with maximum energy appears at $z = 284\lambda$. These protons are found along the propagating axis with $\left|x\right|\lesssim \lambda$ in Fig.~\ref{fig:6}(b). 
The electron bunch in Fig.~\ref{fig:6}(b) around the energetic proton beam corresponds to the electron filament in Figs.~\ref{fig:2}(a) and \ref{fig:4}(a). It also expands transversely in to the region $x\lesssim 10 \lambda$, where magnetic vortex evanescent. According Eq. \eqref{ext}, the on-axis acceleration field $E_z$ is weakened when the negative radial gradient of the magnetic vortex moves radially from the propagating axis, as shown by the solid green curves in Figs.~\ref{fig:6}(c) and (d).

\section{Results and discussion}

\begin{figure}[h]
\includegraphics[width=0.45\textwidth]{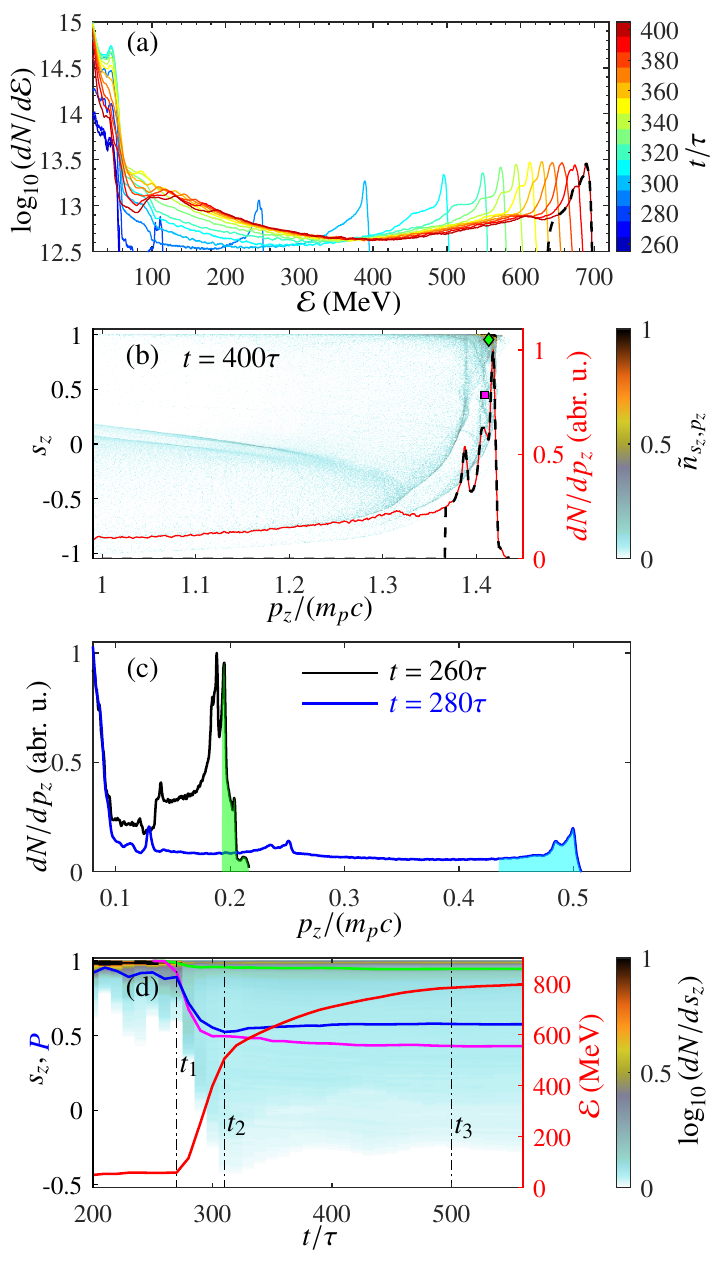}
\caption{\label{fig:7} (a)  Energy spectra of protons at different time during $260\tau \leq t \leq 400\tau$. (b) Density plot of protons in space of $s_z$ and $p_z$ at $t=400 \tau$. The spectrum on $p_z$ for all protons and beam protons are given by the solid red and dashed black lines, respectively. (c) The spectrum on $p_z$ at $t=260\tau$ and $280\tau$ are given by the black and blue lines, respectively. The shaded areas indicate the protons corresponding to those in the final accelerated beam.  (d) The spectrum of beam protons on longitudinal component of the spin vector $dN/ds_z$, as a function of time. The solid blue line describes the dependence of the beam polarization $P=\left|\left<\vec s\right>\right|$ on time.The cutoff energy of the accelerated protons is plotted with the dashed red line. The dashed black line in (a) corresponds to the accelerated beam protons shown with the black dashed line in (b). The magenta cube and the green diamond in (b) denote the locations of the two traced particles, respectively, same as those in Fig. \ref{fig:4}. The magenta and green curves in (d) show their time dependent $s_z$, respectively. } 
\end{figure}

When the energetic protons in the plasma wave are trapped by  the magnetic vortex enhanced acceleration field, a mono-energetic proton beam can be produced.
Energy spectra of protons from $t=260\tau$ to $400\tau$ are plotted with curves of different colors in Fig.~\ref{fig:7}(a). 
When the laser pulse propagates in the gas jet $t\leq 270 \tau$, the cutoff energy is around $61 $~MeV, corresponding to the radially accelerated proton energy by the plasma channel $\mathcal{E}_\mathrm{ch}$. The cutoff energy is almost doubled to 116~MeV at $t=280 \tau$. 
Although the lower boundary of the MVA field in Figs.~\ref{fig:4}(b) and \ref{fig:6}(c) stays at $z_2$,  its amplitude reduces significantly when the negative radial gradient leaves radially from the propagating axis, as discussed in Figs.~\ref{fig:6}(c) and (d). Therefore,  only a portion of the protons in the plasma wave with higher longitudinal velocity component $p_z$, shown in Figs.~\ref{fig:2}(e) and \ref{fig:5}(a),  are trapped.
As a result, energy peaks are found in spectra at $t \geq 280\tau$ in Fig.~\ref{fig:7}(a). 
With the continuous acceleration of the trapped beam, the peak energy reaches 689~MeV at $t=400\tau$.
The production of the energy peak stems from the pre-acceleration of protons within the bubble, driven by a PW laser with an optimized plasma density. On the one hand, the effect of BA may diminish in plasmas with higher density, resulting in an energy spectrum of MVA protons that lacks a distinct energy peak~\cite{Nakamura:2010:PhysRevLett.105.135002,Park:2019:POP,Tazes:SR:2024}. On the other hand, self-focusing of the laser beam is weakened in plasmas with lower density, and the reduction of $a_0$ in Eqs.~\eqref{eq:ezm} and \eqref{eq:pzm} decreases the acceleration of the protons.

\begin{figure}[t]
\includegraphics[width=0.45\textwidth]{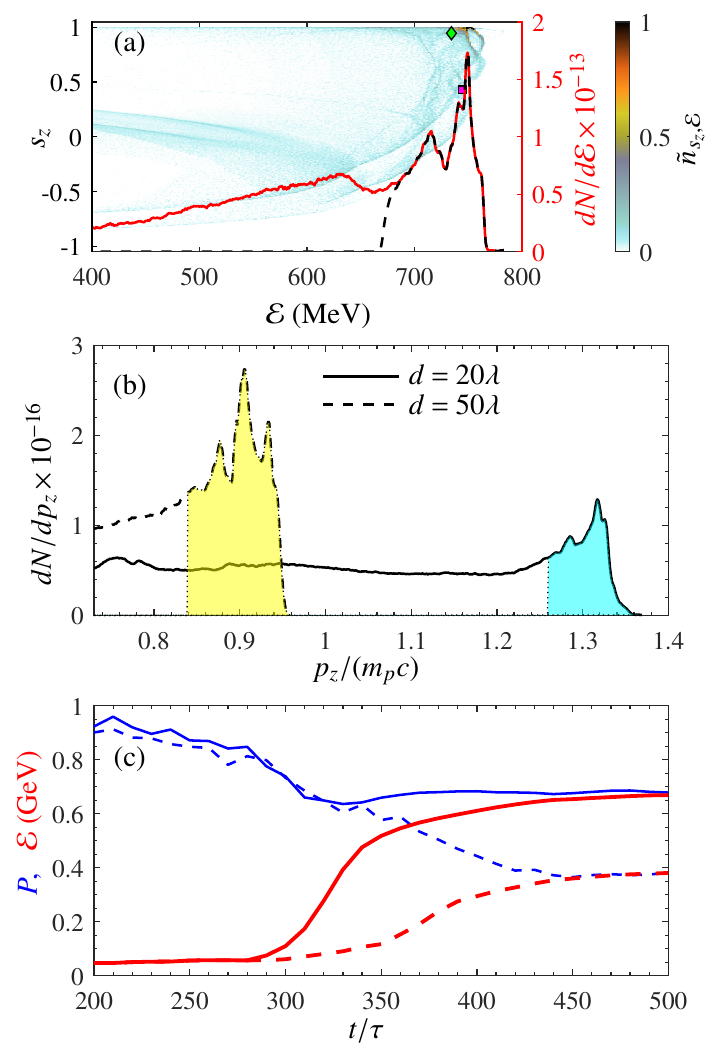}
\caption{\label{fig:8}  (a) Density plot of protons at $t=500 \tau$ in the space of $s_z$ and particle energy $\mathcal{E}$, as well as the spectrum for all energetic protons and the accelerated beam protons, represented by the solid red line and the dashed black line, respectively. The magenta cube and the green diamond denote the locations of the two traced particles, respectively, as in Fig. \ref{fig:4}. 
Longitudinal momentum spectra of accelerated protons at $t=500\tau$ (b), and the evolution of maximum energy and beam polarization over time (c) for cases with different density ramp lengths $d$. In (b) and (c), solid and dashed curves represent cases with $d = 20\lambda$ and $50 \lambda$, respectively. } 
\end{figure}

The spin polarization of the accelerated proton beam is given by $P=\left|\left<\vec s\right>\right|$, where $\left<\right>$ indicates the average over all protons in the beam. In our setup, it is roughly determined by $\left<s_z\right>$ when the spin precession of beam protons occurs in the cylindrically symmetric magnetic vortex $B_\varphi$. The beam polarization is calculated on the basis of identifying beam protons, a. k. a. the protons in the accelerated beam.
Since there is no energy peak in the spectra at $t <280 \tau$  in Fig.~\ref{fig:7}(a), beam protons can not be identified with the energy spectrum when the laser pulse propagates in the gas jet. Alternatively, we identify the proton beam with protons in the peak of the longitudinal momentum spectrum on $p_z$, with $p_z > p_{z0}$. This is reasonable because, as shown in Fig. 5, both before and after the injection of MVA, the protons in the accelerated beam have relatively high $p_z$. 
The density plot of all energetic protons in space along $s_z$ and $p_z$ at $t=400\tau$ is shown in Fig.~\ref{fig:7}(b), where the longitudinal momentum spectra of all protons and the beam protons are plotted with solid red and dashed black curves, respectively. The lower limit of the longitudinal momentum of the proton beam is given by $p_{z0} = 95\% p_{z,\mathrm{peak}}$, where $ p_{z,\mathrm{peak}} = 1.42 m_p c$ corresponds to the peak value of the momentum spectrum at $t=400\tau$.
The quantity of beam protons can be calculated in the momentum spectrum as 
\begin{align}\label{n0}
N_0=\int_{p_{z0}}^\infty (dN/dp_z) dp_z \,.
\end{align}   
Since the number of protons in the accelerated beam remains stable, one can consider protons with relatively high momentum, in the same quantity $N_0$, as representative of the beam protons throughout the entire acceleration process. In other words, at times $t \neq 400 \tau$, the lower limit of momentum $p_{z0}$ is determined by Eq.~\eqref{n0} with $N_0=N_0|_{t = 400 \tau}$, rather than $p_{z,\mathrm{peak}}$. The beam protons at $t=260\tau$ and $280\tau$ are depicted with the shaded areas of the momentum spectra in Fig.~\ref{fig:7}(c). It reveals that only a small fraction of protons at the momentum peak at $t=260\tau$, while still within the target, will be trapped into the final accelerated beam.
 The corresponding energy spectrum of the beam protons is plotted with dashed black line in Figs.~\ref{fig:7}(a) and \ref{fig:8}(a). 
With the definition of beam protons, protons of same quantity $N_0$ with $p_z \geq p_{z0}$ are taken into account to calculate the beam polarization in the whole process of acceleration. The beam polarization $\left<s_z\right>$ is thus obtained, with $P=92\%,\, 55\%$ and $57\%$ at $t=200\tau,\,300\tau$ and $400\tau$, respectively. 

The time evolution of beam polarization is shown by the blue solid curve in Fig.~\ref{fig:7}(d), where the red solid line presenting the evolution of the cutoff energy of the accelerated proton beam. It shows the protons acquire energy $a_0mc^2/2$ before $t_1=270\tau$, when the laser pulse propagates in the plasma. As presented in Fig.~\ref{fig:2}(f) and \ref{fig:3}(b), the proton beam consists of
protons with relatively high longitudinal momentum. Therefore the beam keeps relatively high polarization when the laser propagates inside the target. 
Strong energy increment and strong depolarization in Fig.~\ref{fig:7}(d) appear in the time duration $t_1 \leq t \leq t_2$, when the laser passes though the rear boundary of the target and energetic protons experiences MVA. The MVA almost ends when the energy reaches 495~MeV at $t_2=310 \tau$. 
Subsequently, the acceleration field around the propagating axis becomes weaker, and the magnetic field almost vanishes, as seen in Figs.~\ref{fig:6}(c) and (d). The protons are accelerated slowly with the residual electron bunch till $t_3=500\tau$, and the proton beam polarization sustains after $t_2$. 
The density plot of protons in the space of $s_z$ and particle energy $\mathcal{E}$ at $t=500 \tau$ is depicted in Fig.~\ref{fig:8}(a). The solid red curve and the dashed black curve represent the energy spectra of all protons and the beam protons, as defined by Eq.~\eqref{n0}, respectively. The energy spectrum reveals the energy peak at 749~MeV, and the beam polarization is $66\%$. 


In order to compare spin dynamics of different protons in the accelerated beams, we trace two typical protons with the same momentum $p_z = 1.41 m_p c$ but different $s_z$ in the momentum peak of Fig.~\ref{fig:7}(b). They are named as proton A and B, as well as marked by green diamond and magenta cube, respectively. Corresponding evolutions of their $s_z$ are plotted by green and magenta solid lines in Fig.~\ref{fig:7}(d), where the longitudinal spin component of proton A stays $>94.5\%$ for the whole progress, and for proton B one finds $s_z<50\%$ when $t>t_2$. The different spin dynamics of protons A and B can be understood from their trajectories, which are plotted by green and magenta curves in Figs.~\ref{fig:4}(a) and \ref{fig:6}(b). It is found that both of them originate from the down ramp of the gas jet, with $z_2\lesssim z<z_2+d$. More importantly, proton A originates from the propagating axis of the drive pulse and moves along it. Differently, proton B originates from an off-axis location and moves forward with a divergence angle. When protons A and B are accelerated in the almost same electric fields, A stays along with propagating axis with zero magnetic field, and B experiences stronger magnetic field beside the propagating axis,  as shown by the locations marked with green diamond and magenta cube in Figs.~\ref{fig:4}(a)-(c) and \ref{fig:6}(b)-(d). The positions of protons A and B in the space of $s_z$ and particle energy $\mathcal{E}$ at $t=500 \tau$ are marked with green diamond and magenta cube in Fig.~\ref{fig:8}(a), respectively. One may conclude that protons with initial smaller deviation from the propagating axis contributes to the portion with higher polarization in the accelerated beam.

Longitudinal momentum spectra of accelerated proton beams at $t=500\tau$ from gas jets with $d=20\lambda$ and $50 \lambda$ are shown in Fig.~\ref{fig:8}(b), where the peaks appears at $p_z=1.32 m_p c$ and $0.91m_p c$, respectively. If the beam protons are defined with the same lower limit of the proton momentum $p_{z0} = 95\% p_{z,\mathrm{peak}}$ in Fig.~\ref{fig:8}(b), one obtains the beam polarization $67.8\%$ and $38\%$, respectively.
Figure~\ref{fig:8}(c) shows corresponding evolution of polarization of the accelerated beams with blue curves, as well as the cutoff energy with red curves, where solid and dashed curves correspond to cases with $d=20\lambda$ and $50 \lambda$, respectively. In comparison with the results with $d=5\lambda$ shown in Fig.~\ref{fig:7}(c), similar characters are found: the proton beam are intensely accelerated and depolarized when the laser pulse propagates through the density down-ramp. with the corresponding time duration for beam acceleration and depolarization $t_2-t_1$ depends linearly on $d$.
The cutoff energy in Fig.~\ref{fig:8}(c)  is 668~MeV and 380~MeV, respectively. One may find the proton beam obtained with $d=20\lambda$ has similar beam energy and polarization to the case in Fig.~\ref{fig:7}. Therefore, density scale length of the gas jet in tens of micrometers is required within the experimentally realizable gas target
with sharp gradient~\cite{Semushin:RSI:2001, Schmid:RSI:2012, Lorenz:2019:MRE, Engin:PPCF:2019,Zhou:POP:2021,Lei:HPLSE:2023,Lecz:2023:PhysRevResearch.5.023169}.

\begin{figure}[t]
\includegraphics[width=0.45\textwidth]{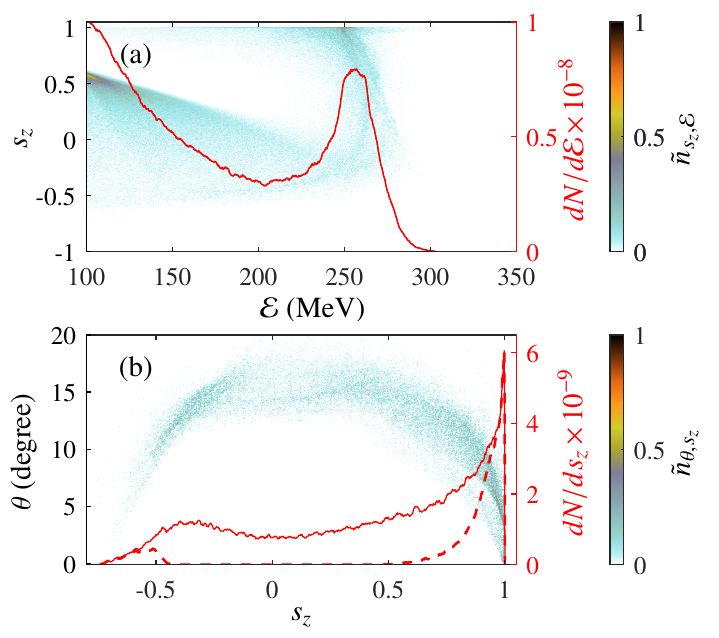}
\caption{\label{fig:9}
(a) Density plot of protons from the three-dimensional simulation at $t=500 \tau$ in the space of $s_z$ and particle energy $\mathcal{E}$, as well as the spectrum for all energetic protons  represented by the solid red line. (b) Density plot of the accelerated protons within the FWHM of the energy peak in panel (a), represented in the space of particle energy $\mathcal{E}$ and spread angle $\theta = \tan^{-1}\left(\left.\sqrt{p_x^2+p_y^2}\right/p_z\right)$. The solid red line indicates the corresponding spectrum along $s_z$. The dashed red line represents the spectrum of protons within the FWHM of the energy peak for those with $\theta<10^\circ$.} 
\end{figure}

 To provide a clear illustration of the acceleration scheme, two-dimensional PIC simulations have been utilized to provide detailed insights into the simulation process. In this context, a three-dimensional PIC simulation has been conducted using the extended EPOCH code to verify the practical efficiency of the acceleration scheme. The size of the simulation box is $180\lambda\times 24\lambda\times 24\lambda$, represented with cells of $1800\times 240\times 240$ and 6 pseudo-particles per cell for protons, electrons and chlorine nuclei. The driving laser's power has been increased to 7.7 PW, with a laser amplitude of $a_0=200$, and a pulse duration of $8\tau$. The plasma density and the length of the uniform plateau are $n_e = 0.72 n_c$ and $z_2-z_1 = 250\lambda$, respectively. The remaining parameters are identical to those used in the two-dimensional simulations for Fig.~\ref{fig:7}.
 
Figure~\ref{fig:9}(a) depicts the energy spectrum of the protons from three-dimensional simulations at $t=550\tau$, and the corresponding density plot in the space of $s_z$ and the proton energy $\mathcal{E}$. The energy peak is observed at 257 MeV. It is notable that the protons at the energy peak have a relatively large $s_z$.
The beam energy is significantly lower than that predicted by two-dimensional simulations, due to the magnetic vortex-induced acceleration field being reduced in the three-dimensional scenario, potentially attributed to the dimensional impact on laser-induced electron heating~\cite{Stark:2017:POP}. Nevertheless, protons are still accelerated to approximately 20 MeV by the laser-driven bubble according Eqs.~\eqref{eq:ezm} and \eqref{eq:pzm}. Their energy is significantly boosted via MVA at the rear boundary of the target.
 Additionally, this polarized proton beams with energy $\gtrsim 200$ MeV can be considered as a direct alternative injector for polarized colliders~\cite{Alekseev:2003,MartinezMarin:2019odq,Mustapha_2020}.

We present the density plot of protons within the full width at half maximum (FWHM) of the energy peak in the $s_z$ and spread angle $\theta$ space in Fig.~\ref{fig:9}(b). The corresponding spectrum on $s_z$ is depicted by the solid red line, indicating a beam polarization of 39.5~\%. The majority of protons are within a spread angle $\theta\lesssim 15^\circ$. The density plot reveals that protons with smaller $\theta$ have higher $s_z$. 
If protons with $\theta < 10^\circ$ are selected, the beam polarization increases to 73.3~\%. The corresponding spectrum of protons with $\theta < 10^\circ$ on $s_z$ is represented by the dashed red line in Fig.~\ref{fig:9}(b). The beam polarization increases further to 88.8~\% if only protons with $\theta < 5^\circ$ are considered. Therefore, higher requirements for beam polarization can be met by using a smaller collection angle. The beam charge at the FWHM of the energy peak is 349 pC. This reduces to 95.4 pC and 23.1 pC in a collection angle of $10^\circ$ and $5^\circ$, respectively.

\section{Conclusion}

Laser driven ion acceleration in a pre-polarized hydrogen halides gaseous target is expected to generate energetic spin polarized protons economically.
When a PW laser passes through a hydrogen halides gas jet, protons are accelerated by the laser driven relativistic channel and wakefield inside the target, as well as the stationary electric field at the rear boundary of the target. Especially, longitudinal motion of protons in the plasma wave triggers the injection of protons into the magnetic vortex enhanced acceleration field.
The acceleration and spin precession of protons is analyzed based on the TBMT equation and analytical models of the laser-driven blowout bubble-channel structure.
It is found that the magnetic vortex  affects the precession of protons in the bubble-channel scheme and the rear boundary of the target. By analyzing dynamics of protons and the structure of the fields, dependence of the final beam energy and polarization on laser and plasma parameters is discussed.
The energy gain is determined by the pre-acceleation of the protons in the plasma wave and the post acceleration out of the target. 
Compared to our initial demonstration of energetic polarized proton beam generation~\cite{Jin:2020:PhysRevE.102.011201}, the proton pre-acceleration within the plasma wave has been enhanced using our analytical model~\cite{Liu:2016:PhysRevAccelBeams.19.073401,Liu:POP:2018}, resulting in a significant increase in the final energy gain.
Results from multidimensional PIC simulations indicate that monoenergetic proton beams with energies in the several hundred MeV range and a polarization of tens of percent can be achieved using petawatt-scale lasers and a pre-polarized gaseous target.
It can serve as an alternative to the polarized ion inject linac of a polarized collider~\cite{Zhang:2019,Alekseev:2003,MartinezMarin:2019odq,Mustapha_2020}.
Our study provides a novel approach to high-energy laser-driven polarized proton acceleration, offering practical proposals for polarized particle colliders.

\begin{acknowledgments}
This work was supported by the National Natural Science Foundation of China (Grant No. 12075081) and the innovation group project of the Natural Science Foundation of Hubei Province of China (Grant No. 2024AFA038). Bin Liu acknowledges the support of Guangdong High Level Innovation Research Institute Project, Grant No. 2021B0909050006.
\end{acknowledgments}

\section*{AUTHOR DECLARATIONS}

\subsection*{Conflict of Interest}
The authors have no conflicts to disclose.

\subsection*{Author Contributions}
\textbf{Zhikun Zou}: Data curation (equal); Software (equal). \textbf{Gan Guo}: Data curation (equal); Validation (equal).
\textbf{Meng Wen}: Funding acquisition (equal); Supervision (equal); Writing - original draft (equal).
\textbf{Bin Liu}: Formal analysis (equal); Funding acquisition (equal); Methodology (equal); Writing - review \& editing (equal).
\textbf{Xue Yan}: Formal analysis (equal); Resources (equal).
\textbf{Yangji\'e Liu}: Conceptualization (equal); Writing - review \& editing  (equal).
\textbf{Luling Jin}: Conceptualization (equal); Supervision(equal); Visualization(equal); Writing - review \& editing  (equal).

\section*{Data Availability Statement}

The data that support the findings of this study are available from the corresponding author upon reasonable request.

\bibliography{bibliography}

\end{document}